\documentclass[aps,prl,twocolumn,letterpaper]{revtex4}
\usepackage{graphicx}
\usepackage{dcolumn}
\usepackage{amsmath}
\usepackage{amssymb}
\usepackage{float}
\usepackage{epstopdf}
\usepackage{color}

\def\kv{{\bf k}}

\def\beq{\begin{equation}}
\def\eeq{\end{equation}}
\def\beqa{\begin{eqnarray}}
\def\eeqa{\end{eqnarray}}

\begin{document}

\title{Edge State Transport in Floquet Topological Insulators}
\author{Aaron Farrell and T. Pereg-Barnea}
\affiliation{Department of Physics and the Centre for Physics of Materials, McGill University, Montreal, Quebec,
Canada H3A 2T8}
\date{\today}
\begin{abstract}
Floquet topological insulators are systems in which the topology emerges out of equilibrium when a time periodic perturbation is applied.  In these systems one can define quasi-energy states which replace the quilibrium stationary states.  The system exhibits its non-trivial topology by developing edge localized quasi-energy states which lie in a gap of the quasi energy spectrum.
These states represent a non-equilibrium analogue of the topologically protected edge-states in equilibrium topological insulators.  In equilibrium these edge-states lead to very specific transport properties, in particular the two-terminal conductivity of these systems is $2e^2/h$. Here we explore the transport properties of the edge-states in a Floquet topological insulator. In stark contrast to the equilibrium result, we find that the two terminal conductivity of these edge states is significantly different from $2e^2/h$. This fact notwithstanding, we find that for certain external potential strengths the conductivity is smaller than $2e^2/h$ and robust to the effects of disorder and smooth changes to the Hamiltonian's parameters. This robustness is reminiscent of the robustness found in equilibrium topological insulators. We provide an intuitive understanding of the reduction of the conductivity in terms of scattering by photons. This leads us to consider a previously proposed Floquet sum rule\cite{Kundu} which recovers the equilibrium value of  $2e^2/h$ for the conductivity when edge states are present. We show that this sum rule holds in our system using both numerical and analytic techniques.
\end{abstract}
\maketitle

\emph{Introduction.}--- Over the past decade topological insulators have become well known for their novel transport properties. The hallmark of these systems is their linearly dispersing, in-gap states. These states correspond to counter-propagating, helical edge modes. In a two dimensional geometry these edge modes represent one dimensional channels and lead to specific transport properties.

One example is a two-terminal device, where a source and drain are attached to the left and right of a sample and a bias voltage is applied across these terminals. The conductivity for the Fermi energy placed in the gap (where the edge states live) is  $\sigma= 2e^2/h$\cite{Bernevig, Roth, konig}. In a six-terminal, or Hall-bar, geometry specific values of multi-terminal resistances are expected\cite{Bernevig, Roth, konig} and these resistances are unique to counter-propagating, helical edge modes.

Although the number of confirmed topological insulators is ever increasing, materials with the correct physical parameters to support this state of matter are hard to come by. This has led many authors to consider ways in which to drive a material without any topological properties into a topological state. When a time-periodic potential is used to accomplish this task the resulting non-equilibrium topological state is called a Floquet topological insulator.

The field of Floquet topological insulators (and Floquet topological superconductors) has produced many interesting results of late\cite{Lindner, Gu,Oka,Usaj, Calvo,Torres,Wang, Leon, Rudner, Kitagawa2,Kundu2,Katan, Jiang, Kundu, Liu, Wu, Wang2, Delplace, Li4}. The introduction of a time-periodic potential into the system breaks continuous time-translational invariance and so one must dispense with the notion of an energy spectrum. A time-periodic field does have discrete time-translational invariance and therefore one has the ability to define an analogous concept called the ``quasi-energy" spectrum\cite{Sambe}. In Floquet topological insulators one uses an externally applied time-periodic field of carefully chosen parameters to drive the system into its topological phase.  The topology is manifest in linearly dispersing, in-gap, edge modes can be created in the quasi-energy spectrum. Such a system then represents a non-equilibrium analogue of topological insulators, but with the added flexibility of am external periodic potential.

In this Letter we have studied the transport properties of Floquet edge-states.
Our goal is to test weather transport through the edge modes of a two dimensional Floquet topological insulator is quantized and robust as in the case of equilibrium topological insulators.
We focus on a two-terminal geometry, although our results and the intuition developed here readily generalizes to other geometries. In general we find that the two-terminal conductivity of the Floquet edge-states is significantly different from the typical equilibrium value of $2e^2/h$ and can be either larger or smaller than this distinctive value depending on how the strength of the external field is tuned.

The main results of this Letter may be summarized as follows.
The existence of quasi-energy edge states in the Floquet topological insulator is accompanied by a conductivity of $\sigma<2e^2/h$, when the chemical potential lies in the quasi-energy gap.  In addition, the value of $2e^2/h$ is obtained as a sum rule when the conductivity is summed over all `side bands', i.e, over all energies which differ from a particular energy in the gap by an integer number of photon energies. Physically, the result $\sigma<2e^2/h$ for non-equilibrium edge states corresponds to the presence of photons inhibiting access to the topologically protected edge states of the system.

Moreover, in regions where the conductivity is smaller than $2e^2/h$, we find that the calculated values are robust to the effects of disorder, system size and (smooth) changes to the system Hamiltonian. Such behavior is reminiscent of topologically protected edge states in equilibrium topological insulators and we indeed find that for the external potential strengths where we see this robustness there exist linearly dispersing, in-gap edge states in the quasi-energy spectrum. In regions where the conductivity is larger than $2e^2/h$ no robustness exists and the gap is closed, hence we are probing bulk effects.

The reduction of the topologically protected conductivity away from $2e^2/h$ can be intuitively understood by borrowing some machinery from the field of photon assisted tunneling (PAT). Namely, an electron that would normally tunnel into the edge-states of the system has a finite probability of absorbing/emitting a photon and being scattered out of the edge state. From the viewpoint of quasi-energy states this comes from understanding that the definite energy states of the leads do not perfectly overlap with the quasi-energy states of the Floquet topological insulator\cite{Kundu2}. The heuristic description in terms of scattering of electrons by photons can be applied to observe a so-called ``Floquet sum rule"\cite{Kundu}. In short, the sum rule recovers all of the conductivity lost from PAT by summing over lead energies separated by photon energies $\hbar \Omega$, $\Omega$ being the frequency of the driving field. We have confirmed this sum rule using both numerics and an approximate analytic approach.

\emph{Model.}--- Our model Hamiltonian is that of a quantum well heterostructure\cite{Bernevig} irradiated by linearly polarized light and subjected to a disorder potential. It is given as follows
\beqa
H_S=\sum_{\kv} \psi_\kv^\dagger \left(   \begin{matrix}   \hat{ H}(\kv, t) & 0 \\   0 &  \hat{ H}^*(-\kv, t)\\ \end{matrix}\right) \psi_\kv-\sum_{i,\alpha} w_i \psi^\dagger_{i,\alpha} \psi_{i,\alpha}
\eeqa
where $\psi_\kv^\dagger$ is a four component creation operator for electrons at momenta $\kv$ in angular momentum state $m_J=(1/2,3/2,-1/2,-3/2)$ and $ \psi_i^\dagger$ is its Fourier transform. The first term above is the Hamiltonian of the clean, irradiated heterostructure and we have used $\hat{ H}(\kv)= \epsilon_\kv \sigma_0 + {\bf d}(\kv)\cdot {\bf \sigma}+2 ({\bf V}\cdot \sigma) \cos{\Omega t}$. The second term takes into account disorder. We have used the standard definitions ${\bf d}(\kv)=(A\sin{k_x},A\sin{k_y}, M-4B+2B(\cos{k_x}+\cos{k_y}))$ and $\epsilon_\kv =C-2D(2-\cos{k_x}-\cos{k_y})$ and draw the $\{w_i\}$ randomly from an evenly distributed sample between $-W/2$ and $W/2$.

Following Lindner and coworkers\cite{Lindner}, we set $C=D=0$, $A=B=0.2|M|$ and set $|M|=1$ throughout ({\it i.e.} all energies are in units of $|M|$). To simulate a trivial system we set $M=-1$ so that $\text{sgn}(M/B)=-1$\cite{Lindner,Bernevig}.  Furthermore, we take ${\bf V}=V_{ext}\hat{z}$ for concreteness. The system outlined above is similar to the system studied in Ref.~\cite{Farrelltop} with the important distinction that here we focus on the trivial region of parameter space. We also note that in this system the existence of the Floquet topological insulating state relies crucially on band-mixing that comes from the periodic perturbation being {\em on-resonance}\cite{Lindner}, i.e. the quantity $\hbar \Omega$ connects different parts of the band structure. This is in contrast to other systems, for example graphene\cite{Gu,Oka,Usaj, Calvo,Torres, Kitagawa}, where the Floquet topological insulator can be driven using an off-resonant perturbation\cite{Kitagawa}.

Our understanding relies primarily on Floquet states\cite{Sambe}. Floquet states are the extension of stationary states to time-periodic systems.  In a time-periodic system one deals with (Floquet) states that solve the Schr\"odinger equation and are characterized by a definite quasi-energy. These states are traditionally written as $|\psi_{\tilde{\eta}}(t)\rangle = e^{-i\tilde{\eta} t/\hbar}|\phi_{\tilde{\eta}}(t)\rangle$, which leads to the eigenvalue equation $\left(H(t)-i\hbar \partial_t\right)|\phi_{\tilde{\eta}}(t)\rangle =\tilde{\eta} |\phi_{\tilde{\eta}}(t)\rangle $ where $H(t)$ is the full Hamiltonian of the system, $\tilde{\eta}$ are the quasi-energies and $|\phi_{\tilde{\eta}}(t+T)\rangle=|\phi_{\tilde{\eta}}(t)\rangle$. We note that if $|\phi_{\tilde{\eta}}(t)\rangle$ is an eigenstate with quasi-energy $\tilde{\eta}$, then $e^{i\Omega t}  |\phi_{\tilde{\eta}}(t)\rangle$ is also an eigenstate but with quasi-energy $\tilde{\eta}+\hbar \Omega$. Therefore the quasi-energy spectrum is only unique up to integer multiples of $\hbar\Omega$. his allows us to define a ``Brillouin zone" for the quasi-energies, we will call this the Floquet zone. For this work we consider $0\le{\eta}<\hbar\Omega$, we will use the convention ${\eta}$ to denote quasi-energies confined to this zone. This reflects the fact that energy in a time periodic system is only conserved modulo $\hbar\Omega$, an electron in a quasi-energy state $|\phi_{{\eta}}(t)\rangle$ can always absorb or emit a photon.

\begin{figure}[]
  \setlength{\unitlength}{1mm}
     \includegraphics[scale=.4]{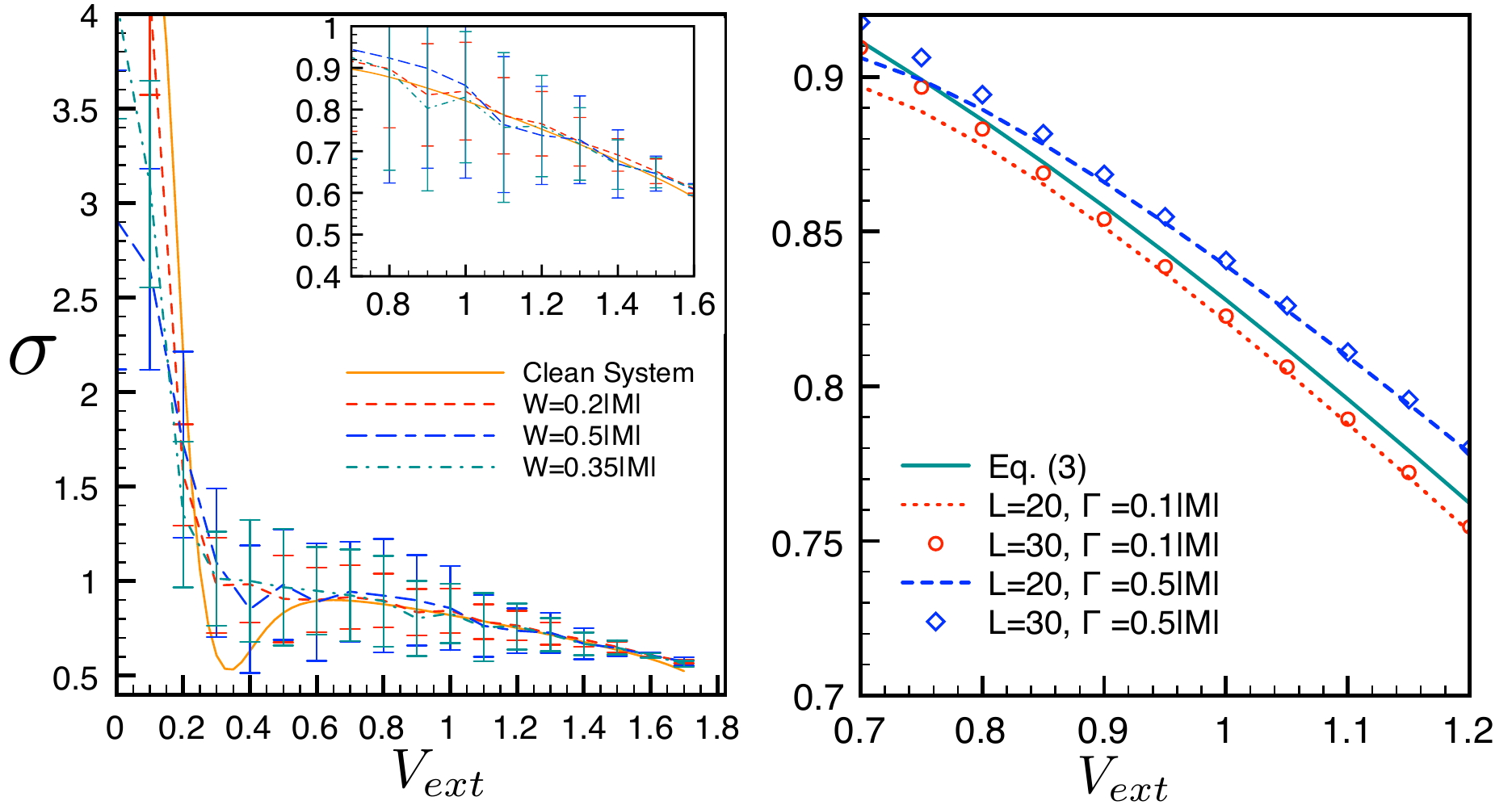}

\caption{{\small
Two terminal conductivity for different external potential strengths in units of $e^2/h$. The left shows the conductivity at $E=\Omega/2$ in a two terminal set up as a function of external potential strength for various disorder strengths.  On the right is the conductivity for various values of the system size, $L$, and the lead coupling parameter $\Gamma$. }
     }\label{fig:TrivSystem}
\end{figure}

\emph{Two Terminal Conductivity.}--- Let us begin with our results for the two-terminal conductivity of this system.  We calculate the conductivity numerically using Floquet-Landauer theory\cite{Torres, Kitagawa}. In this two-terminal setup we consider the leads to be kept at a voltage such that the Fermi level of both leads, which we will refer to as the lead energy, takes a value $E$. We study the differential conductivity at a lead energy of $E=\Omega/2$ which is where the edge states are expected to be found\cite{Lindner}. Referring to our results in Fig.~\ref{fig:TrivSystem}a we see that, with the exception of a small area near $V_{ext}=0.3|M|$, the two-terminal conductivity generally decreases with $V_{ext}$ in the range of parameters considered. We note that nowhere do we see a saturation to a value of $\sigma=2e^2/h$, nor any other constant value. This fact notwithstanding, our results do have the striking feature that after a certain value of $V_{ext}$ the conductivity becomes insensitive to the effects of disorder; in that region all of the curves overlap. In Fig.~\ref{fig:TrivSystem}b we see that in this same region our results are insensitive to system length $L$ and to the parameter $\Gamma$ which describes the strength of the coupling to the leads. Thus we note our first result, for some values of $V_{ext}$ the calculated conductivity is robust in the same way as for an equilibrium topological insulator.

\begin{figure}[]
  \setlength{\unitlength}{1mm}
\begin{tabular}{c}
   \includegraphics[scale=.3]{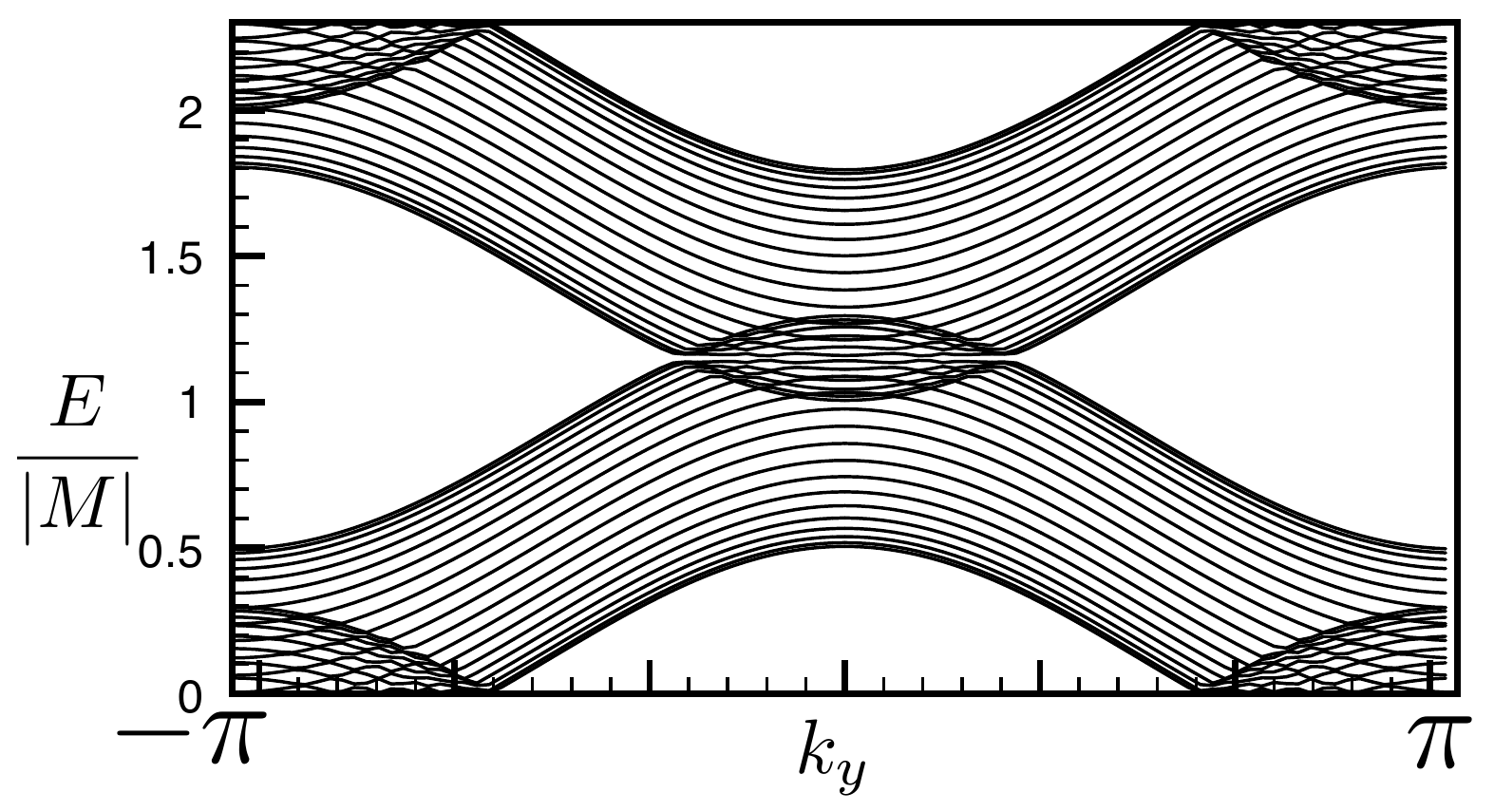} \\
   \includegraphics[scale=.3]{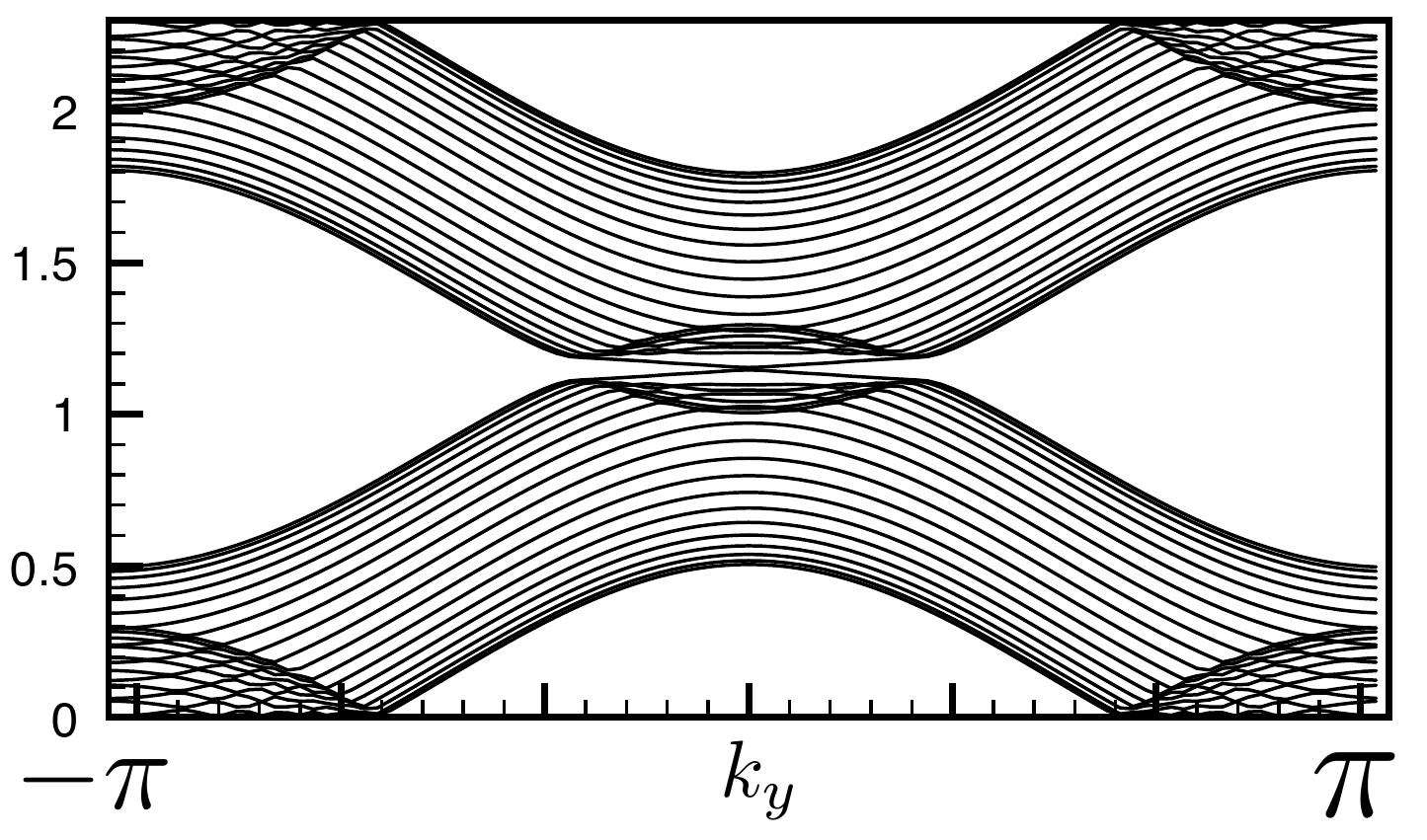} \\
      \includegraphics[scale=.3]{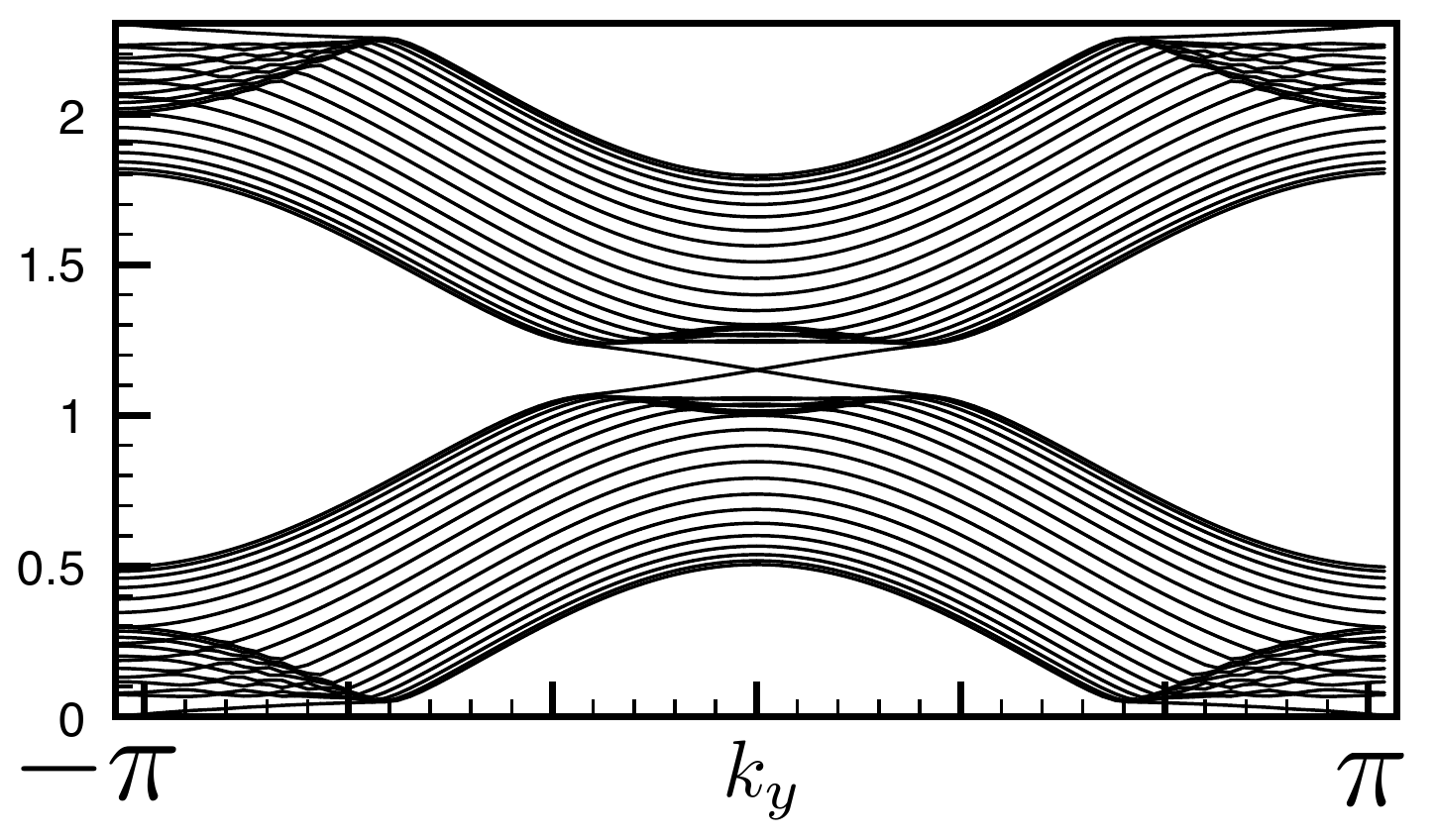} \\
   \end{tabular}

\caption{{\small
Quasi-energy spectrum of a trivial sample at different driving amplitudes. The top plot is for $V_{ext}=0.1|M|$, the middle  $V_{ext}=0.3|M|$, and the bottom  $V_{ext}=0.9|M|$.      }
     }\label{fig:QEM2}
\end{figure}

The robustness in the conductivity coexists with the presence of edge-states in the quasi-energy spectrum. To show this we consider the system in the absence of leads and in a cylindrical geometry. The quasi-energy for our model appears in Fig.~\ref{fig:QEM2} for several values of the external potential strength $V_{ext}$. For small driving strength the gap remains closed, but as the strength is increased the gap opens up leaving linearly dispersing states. Further inspection of these states reveals that they reside on the edge of the system\cite{Lindner}. In general we have found that when this gap is open and large enough to withstand the effects of disorder or coupling to the leads, the value calculated for the conductivity is robust in the same sense as edge states in a topological insulator.

\emph{Photon Inhibited Transport and Floquet Sum Rule.}--- Thus far we have shown that when this system plays host to edge states the conductivity we find appears to be topologically robust. We now address the question of why it does not have the hallmark value of $2e^2/h$. For this we further generalize a technique inspired by photon-assisted tunnelling\cite{TienGordon, Platero20041} and used in Ref. \cite{Farrelltop}. In this work it was shown that for a topological heterostructure the presence of an external time-periodic field reduces the conductivity away from $2e^2/h$. This reduction and other subsequent non-trivial results can be accounted for by understanding that the external potential not only ``dresses" the quantum well Hamiltonian, but also splits this dressed system into side-bands\cite{TienGordon, Platero20041}. The splitting means that the edge states of the system are only populated probabilistically, accounting for the reduction in the standard transport quantities. The specific application of Ref. \cite{Farrelltop} relied crucially on the driving potential being {\em off-resonance}, {\it i.e.} that it did not mix portions of the equilibrium band structure. The current system requires {\em on-resonance} light in order to drive the system into a topological state. In spite of this, our results are conducive to a similar interpretation in that we see topologically robust results in Fig.~\ref{fig:TrivSystem} that are different from $2e^2/h$.

To put this discussion on more general grounds we appeal to Floquet theory. As discussed previously, in a time-periodic system the states of interest are the steady state solutions $|\psi_{\eta}(t)\rangle = e^{-i\eta t/\hbar} |\phi_{\eta}(t)\rangle$, where $\eta$ is the quasi-energy and $\left(H(t)-i\hbar \partial_t\right)|\phi_{\eta}(t)\rangle =\eta |\phi_{\eta}(t)\rangle $. For periodic dependence in $t$ we are free to define the following decomposition
\beq\label{sidebands}
 |\phi_{\eta}(t)\rangle = \sum_{n} e^{-in\Omega t}  |\phi_n\rangle
\eeq
In the literature the states $ |\phi_n\rangle $ are commonly called sideband states\cite{Kohler} and are determined as solutions to the eigenvalue equation $\sum_n \bar{H}_{n,m} |\phi_m\rangle = (\eta+n\hbar\Omega) |\phi_n\rangle$ where $ \bar{H}_{n,m}= \frac{1}{T}\int_0^T dt e^{i(n-m)\Omega t}  H(t)$.

We now calculate the time-averaged expectation value of the energy in the steady state $|\psi_{\eta}(t)\rangle$ which we define as $\bar{E}=\frac{1}{T}\int_0^T dt  \langle \psi_\eta(t)|H(t)|\psi_{\eta}(t)\rangle$. Using  the side-band decomposition in equation (\ref{sidebands}) and the fact that $|\psi_{\eta}(t)\rangle$ solves the Schr\"odingier equation immediately gives
\beq\label{barE}
\bar{E}=\sum_{n}(\eta+\hbar \Omega n) \langle \phi_n|\phi_n\rangle
\eeq
Noting that $ \langle \phi_n|\phi_n\rangle \ge0$ and $\sum_n  \langle \phi_n|\phi_n\rangle =1$ (the latter property follows from the normalization of $ |\phi_{\epsilon}(t)\rangle $) allows us to interpret the above average as follows. In the quantum state $|\psi_{\eta}(t)\rangle$ the energies $\eta+\hbar\Omega n$ occur with probability $\langle \phi_n|\phi_n\rangle $.

We now tie the above statistical interpretation to our observations of the transport in the Floquet topological insulators. For the system of interest one can calculate the appropriate quasi-energies  $0\le \eta < \hbar\Omega$ and their corresponding wave functions $ |\phi_{\eta}(t)\rangle $, these are the steady states of our sample. Now, when electrons from the lead are injected into the system at some definite energy $E$, as opposed to an equilibrium case, only a portion of the sample state overlaps with the definite energy lead state\cite{Kundu2}. Physically, we envision this in terms of electrons being able to absorb or emit photons once they enter the sample. For lead electrons at an energy $E=\eta+N\hbar\Omega$ there is only a probability $\langle \phi_N|\phi_N\rangle$ that the electron will absorb/emit enough photons to access the sample state with quasi energy $\eta$. This quasi-energy spectrum may contain topologically protected edge states\cite{Leon, Rudner, Kitagawa2,Kundu2,Katan}. Now, when we try to access these states from a charge transport point of view we can only access the state  at a certain probability, because of the possibility to absorb/emit photons. Therefore expected signatures of these edge states, {\it e.g.} $\sigma=2e^2/h$ conductance, are probabilistically suppressed.

Note that this argument {\em does not} rely on the periodic perturbation being on or off resonance, it is simply a consequence of the discrete time-translational invariance. Therefore, when one is dealing with Floquet edge-states it should be kept in mind that the weight of these edge states is distributed into sidebands as discussed above. Indeed in the current system one can approximately obtain a description of the conductivity at the specific lead energies $E+n\hbar\Omega$ (where the edge-states live). Quoting only the result here\cite{Supp}
\beqa
\sigma(E+N\hbar\Omega)&\simeq&  \frac{1}{2}\left(J^2_{N}\left(\frac{2V_{ext}}{ \hbar \Omega}\right) +J^2_{N+1}\left(\frac{2V_{ext}}{ \hbar \Omega}\right) \right)\nonumber  \\ &\times&\tilde{\sigma}(E, V_{ext}) 
\eeqa
where the relevant energy $E$ is chosen to be in the vicinity of $\hbar\Omega/2$ where the localized quasi-energy states appear. $\tilde{\sigma}(E, V_{ext})$ is a complicated function of the model parameters and, interestingly, {\em cannot} be thought of as the conductivity of some effective static system. We find numerically that $\tilde{\sigma}(E, V_{ext})\simeq 2e^2/h$ when edge-states are present in the quasi-energy spectrum. The important implication of the above formula is that the conductivity can be thought of as an overall probabilistic factor times a conductivity of the underlying system.  The above approximate result compares very well to our numerical calculations. A plot of this function appears in Fig.~\ref{fig:TrivSystem}b.

\begin{figure}[tb]
  \setlength{\unitlength}{1mm}

   \includegraphics[scale=.3]{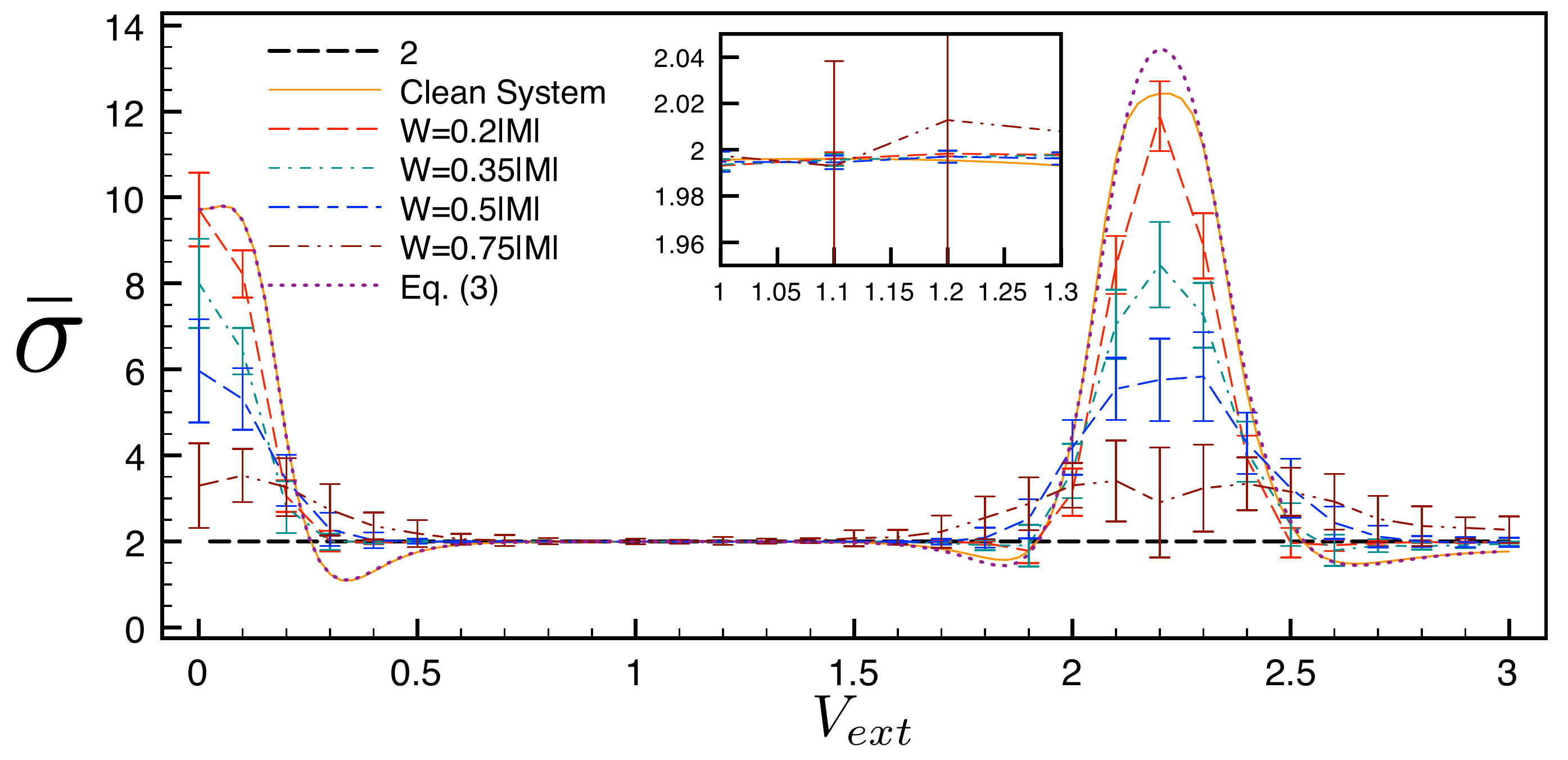}

\caption{{\small
Results for disorder averaged summed conductivity, {\it i.e.} Eq. (\ref{sumrule}) in the text, with $M=-1$ and $E=\Omega/2$ and in units of $e^2/h$.  The inset  shows a zoomed in picture of the first area of conductivity quantization. The disorder plots are constructed by averaging over $40$ randomly drawn collection of disorder potentials while the error bars represent one standard deviation of this data. Note some error bars in the insets are too small to see. This data has been obtained from a calculation on a $20\times20$ lattice.
     }
     }\label{fig:sumruletriv}
\end{figure}

With the intuition for why the conductivity is suppressed in Floquet topological edge-states, let us move on to present results for how the value of $2e^2/h$ can be recovered. In short, by setting lead energies at  $\hbar\Omega/2+n\hbar\Omega$ we should be able to recollect the lost statistical weight from the photon scattering. Towards this end we consider the quantity\cite{Kundu}
\beq\label{sumrule}
\bar{\sigma}(E)=\sum_n \sigma(E+{n\hbar\Omega})=\tilde\sigma(E,V_{ext}).
\eeq
We calculate $\bar{\sigma}(E\simeq \Omega/2)$ for various different values of $V_{ext}$ and also at different disorder strengths. Our results are presented in Fig.~\ref{fig:sumruletriv}. What we see is quite satisfying: for a window of $V_{ext}$ values we see that $\bar{\sigma}=2e^2/h$. Moreover, this window of $V_{ext}$ values corresponds to the same parameter regime where there are in-gap quasi-energy edge states, and insensitivity of the system to disorder, system size and other parameters in Figs.~\ref{fig:TrivSystem}, \ref{fig:QEM2}.

We understand the plot in Fig.~\ref{fig:sumruletriv} as follows. For smaller $V_{ext}$ the external field is not strong enough to open a gap and ``expose" the edge states. Therefore the conduction $\sigma$ is a result of bulk processes and thus sensitive to disorder strength. As $V_{ext}$ gets large enough to open a sufficiently stable gap the edge states appear in this gap and are unobscured by bulk states.  Here we see $\bar{\sigma}=2e^2/h$ and an {\em insensitivity} to disorder strength. Eventually $V_{ext}$ becomes so strong that the gap closes again and bulk states dominate.  In this case $\bar{\sigma}(\hbar\Omega/2) > 2e^2/h$ representing bulk conductivity. When the gap opens again at a larger external potential we see a reversion back to $\bar{\sigma}=2 e^2/h$.

\emph{Conclusions.}---We have explored the transport properties of Floquet topological edge-states in a quantum well heterostructure. At first we took a numerical approach which showed that in the presence of Floquet edge states in the quasi-energy spectrum the two-terminal conductivity is topologically robust, albeit  {\em not} quantized to $2e^2/h$.

To explain the reduction of the two-terminal conductivity compared to the equilibrium value of  $2e^2/h$ we appealed to an intuitive description in terms of electrons being scattered by photons. This picture consists of viewing the Floquet edge states in the quasi-energy spectrum as having their weight distributed into side-bands of energies $\eta+n\hbar\Omega$. The result of this side-band distribution is that as we attempt to inject an electron from a lead at some energy $E$ there is a certain probability that it will absorb/emit enough photons to find the Floquet edge state.

The heuristic picture in terms of scattering by photons motivated us to propose a means to salvage the equilibrium conductivity of $2e^2/h$. This can be done using a recently proposed Floquet sum rule\cite{Kundu}, which in our formalism has a natural interpretation.
In our picture the topological Floquet states represent a superposition of states in various side bands.  The different coefficients in the superposition $\langle\phi_n|\phi_\eta\rangle$ determine the overlap.
Our Floquet edge states nicely obey this sum rule showing a value of $\bar{\sigma}=2e^2/h$ is found when the external field is such that edge states in the quasi-energy spectrum exist. Moreover, the result  $\bar{\sigma}=2e^2/h$ is robust to disorder even up to very large disorder strengths.

The authors are thankful for useful discussions with Jean-Ren\'e Soquet, Aashish Clerk and Gil Refael. Financial support for this work was provided by the NSERC and FQRNT (TPB) and the Vanier Canada Graduate Scholarship (AF). Numerical calculations for this work were performed using McGill HPC supercomputing resources.

\bibliographystyle{apsrev}
\bibliography{Floquet}

\end{document}